\documentstyle[aps]{revtex}

\begin{document}

\title{Strong Discontinuities in the Complex Photonic Band Structure \\
of Transmission Metallic Gratings}

\author{S. Collin, F. Pardo\cite{email}, R. Teissier and J.-L. Pelouard}
\address{Laboratoire de Microstructures et Micro\'electronique, CNRS UPR20,\\
196 av. Henri Ravera, BP 107, 92225 Bagneux CEDEX, France}

\date{July 4, 2000}
\maketitle

\begin{abstract}
Complex photonic band structures (CPBS)
of transmission metallic gratings
with rectangular slits are shown to exhibit strong discontinuities
that are not evidenced in the usual energetic band structures.
These discontinuities are located on Wood's anomalies and reveal
unambiguously two different types of resonances,
which are identified as horizontal and vertical surface-plasmon resonances.
Spectral position and width of peaks in the transmission spectrum
can be directly extracted from CPBS for both kinds of resonances.
\end{abstract}



Recently, metallic films with sub-wavelength apertures
became the subject of increasing interest.
Experiments show that 2D arrays
of sub-wavelength holes in metallic films
can lead to extraordinary transmission of light
\cite{Ebbesen98,Ghaemi98,Thio99,Grupp99,Soennichsen2000}.
These properties have been attributed to the excitation
of coupled surface-plasmons
on the upper and lower surfaces of the grating.
The ability of these structures to control
light has been shown \cite{Kim2000,Thio2000},
and applications have already been proposed
in order to exploit these properties in different fields,
such as electromagnetic filters or photolithography.

Electromagnetic calculations
have been carried out by Porto \emph{et al.} \cite{Porto99}
in order to study the mechanisms
that enhance the transmission
of light through metallic gratings with very narrow slits.
They distinguished two different mechanisms,
that is the excitation of coupled surface plasmon polaritons
on both surfaces of the metallic grating,
and the coupling of incident plane waves
with waveguide resonances located in the slits.
They correspond respectively to surface-plasmon bands
and flat bands in the energetic band structure.
Similar resonances located in grooves have also been observed
experimentally and numerically in the reflected light
of metallic gratings \cite{LopezRios98,Tan99,Wandstall98}.

Complex photonic band structures (CPBS) were previously
calculated by Kuzmiak \emph{et al.} for 2D periodic
systems with metallic components \cite{Kuzmiak97},
in order to obtain the attenuation and the lifetime of each mode.
CPBS allowed a splitting of the lifetime of degenerate modes
at Brillouin-zone boundaries to be observed.

In this Letter, we will show that,
in the case of rectangular metallic gratings,
complex dispersion curves demonstrate
the existence of a new kind of discontinuities.
Much stronger than for 2D periodic systems, they are
located on Wood's anomalies.
They correspond to the transition between two
different types of resonances,
identified as horizontal and vertical surface-plasmon resonances.
The first one is a periodic structure resonance,
the second one is a Fabry-Perot like resonance.
The discontinuity can be up to four orders of magnitude.
Moreover, we will see on zero-order transmission spectra
that the calculated complex frequencies are in excellent accordance
not only with the spectral position,
but also with the width of resonances' peaks.

Fig.~\ref{fig:fig1} shows the structure studied in this Letter,
which is similar to the one of Porto \emph{et al.} \cite{Porto99}.
In the following, the period of the grating $d = 3.5~\mu{\rm{m}}$
and the width of the slits $a = 0.5~\mu{\rm{m}}$ will be kept constant,
whereas the height $h$ of the grating will vary.
It is a symmetric structure, surrounded with air.
The metal is assumed to be gold,
whose dielectric function is taken from
the reference used by Porto \cite{Dold65}.

Our calculations were carried out
for the TM polarization
(the incident magnetic field is parallel to the direction $y$
of the grating, see Fig.~\ref{fig:fig1}).
They are based on the exact modal expansion in region II \cite{Sheng82}
and $S$-matrices formalism.
In the upper and lower regions (I and III),
the electromagnetic field is expressed as
superposition of plane waves
$e^{[i (k_{x}^{(n)} x \pm k_{z}^{(n)} z) - i \omega t]}$
(Rayleigh's expansion).
Their pure real wave vector component $k_{x}^{(n)}$,
which represents the in-plane momentum of photons
is given by Eq.\ (\ref{eq:kpar}):
\begin{equation}
  k_{x}^{(n)} = k_{x}^{(0)} + \frac{2 \pi n}{d}.
  \label{eq:kpar}
\end{equation}
The two other terms $k_{z}^{(n)}$ and $\omega$ are complex quantities,
according to the relation
${\left(k_{x}^{(n)}\right)}^2 + {\left(k_{z}^{(n)}\right)}^2 =
{\left(\omega/c\right)}^2$.

For the calculation of CPBS, we use a $S$-matrix formalism,
in which the amplitude of the reflected and transmitted
diffracted waves are given by a matrix $S$
applied to a vector containing
the amplitude of the upper and lower incident waves.
An electromagnetic mode is defined
by the condition of the existence of a wave
in the structure without any incident wave,
which is equivalent to say
that the inverse ($S^{-1}$) of the $S$-matrix
has an eigenvalue that is null,
with an eigenvector $(R_n,T_n)$, 
corresponding to an electromagnetic field in regions I and III:
 \begin{equation}
   \Psi^{I} = \sum_{n} \: R_n \:
e^{[i (k_{x}^{(n)} x + k_{z}^{(n)} z) - i \omega_0 t]},
   \Psi^{III} = \sum_{n} \: T_n \:
e^{[i (k_{x}^{(n)} x - k_{z}^{(n)} z) - i \omega_0 t]},
 \end{equation}
with $\Re{(i k_{z}^{(n)})} \leq 0 $.

For a given $k_x = k_{x}^{(0)}$,
the electromagnetic modes are defined by complex
frequencies $\omega_0$.
The inverse of the imaginary part of the frequency
provides the amplitude lifetime $2 \tau$ of
the resonance \cite{Pardo86,Kuzmiak97},
which can be associated
to the dimensionless quality factor $Q = \Re(\omega_0) \: \tau$.

The complex dispersion curves are lines
in the three-dimensional space defined by the complex
frequency $\omega_0$ of resonance modes
and the parallel momentum of photons $k_{x}$.
For a more comprehensive representation
we are plotting two projections of $\omega_0$,
the energy
$E = \hbar \: \Re{(\omega_0)}$
and the dimensionless quality factor
$Q = {\Re{(\omega_0)}}/{\left(-2 \: \Im{(\omega_0)}\right)}$.

Fig.~\ref{fig:fig2} shows the complex dispersion curves
calculated for a height $h = 1.4~\mu{\rm{m}}$.
The lines on the energetic band structure
(Fig.~\ref{fig:fig2}(a)) represents the
frontier of the Wood-Rayleigh's anomalies.
They correspond to the emergence (or the vanishing) of
a diffracted wave in the region I or III,
which wave is then at the grazing angle.
These anomalies are observed on transmission spectra when the $n^{\rm th}$
parallel wave vector $k_{x}^{(n)}$ (given by Eq.\ (\ref{eq:kpar}))
is equal to the real incident wave vector $k_0$.

The energetic band structure in Fig.~\ref{fig:fig2}(a) shows
two different types of modes, whose dispersion relations
have already been studied by Porto.
The bands located close to Wood's anomalies,
with slightly smaller energies,
characterize horizontal surface-plasmon resonances.
They are excited by the first non-propagating (or evanescent) diffracted wave,
and their quality factors are greater than 10
and up to more than $10^4$ (see Fig.~\ref{fig:fig2}(b)).
On the other hand, flat bands represent electromagnetic modes
which are almost independent of the incident angle $\theta$,
and with lower quality factors (below 20).
They were previously reported as waveguide resonances \cite{Porto99}.
We will justify further in this Letter why
these resonances are more precisely vertical surface-plasmon resonances.

The most remarkable result is given by the quality factor
of the upper energetic band,
which is broken into three parts.
Considering the point A in Fig.~\ref{fig:fig2}(a),
we see that the energetic band is continuous
whereas the quality factor's curve is discontinuous
at the corresponding points A' and A'' in Fig.~\ref{fig:fig2}(b).
On the left side of point A, the flat band corresponds to
a vertical resonance with very low quality factors (of the order of 3),
whereas the right side corresponds to an horizontal surface-plasmon resonance
with high quality factors (up to more than $10^4$).
The lifetimes of the resonances are $\tau=2.9~{\rm{fs}}$ at A'
and $\tau=33~{\rm{ps}}$ at A''.

In order to understand physically the behavior of the electromagnetic waves,
it must be pointed out that the discontinuous transitions
between the two types of resonance are located exactly
on Wood's anomalies (point A and its symmetrical point B),
so that they are linked to the emergence
or the vanishing of a diffracted wave.
This phenomenon can be explained
in considering the energetic band structure:
below Wood's anomaly, the first non-propagating
(evanescent in $z$ direction) diffracted wave ($n$ = +1 at the point A
and $n$ = -2 at the point B)
is responsible for the excitation
of the horizontal surface-plasmon resonance.
When the resonance crosses Wood's anomaly,
the evanescent wave becomes a $z$-direction propagating diffracted wave
(at grazing angle) which can not excite an horizontal resonance anymore,
but is coupled to vertical surface-plasmons
located on the vertical walls of slits.
Then, we obtain a flat band which corresponds to
a "weaker" resonance in the Fabry-Perot like cavity
delimited by the top and bottom surfaces of the grating.
This vertical resonance is coupled
with the last propagating diffracted wave,
thus it ends at the crossing with Wood's anomaly, where this wave vanishes.

The complex band structure
in Fig.~\ref{fig:fig2} has also a continuous
transition between the horizontal and
the vertical resonances in the lower band.
The flat band at normal incidence corresponds to vertical resonances,
and becomes a surface-plasmon band for greater incident angles.
Its quality factor increases from 20 to about
300 during this smooth transition.
The last flat band at about 0.35~eV corresponds
to a vertical resonance for the whole range of $k_x$
in Fig.~\ref{fig:fig2}.

By varying the width $a$ of the slits,
we have observed that the complex mode
of the vertical resonance is nearly unchanged.
In the calculation, it can be seen from the
modal decomposition of the electromagnetic field,
that this resonance corresponds mainly to the excitation
of the first eigenmode of a $z$ infinitely extended region II.
The wave is evanescent in the $x$ direction and propagates
at the vertical interfaces air/metal ($z$ direction).
This TM mode exists for any width of slit,
and becomes equal to the wave vector $k_{ps}$ of the surface-plasmon of
a semi-infinite air-metal interface in the limit of wide slits,
so that it is more correct to call it a \emph{vertical surface-plasmon mode}
instead of a waveguide mode.
The transitions between the regions II/I and II/III act as
mirrors for the vertical surface-plasmon mode.
Their reflection coefficients are low,
so that the vertical resonance is a Fabry-Perot like
resonance with low quality factors.

Complex dispersion curves shown above can be seen as
convenient compact representations of the resonances,
which can be associated to transmission spectra.
For that purpose, we make the hypothesis that
the complex frequency $\omega_0^i$ of a resonance $i$ is a single pole
of the amplitude of the transmitted waves.
Consequently, the zero-order transmitted energy can be
approximated by a Lorentzian function near the resonance $i$.
Assuming $\omega$ is the (real valued) frequency of the incident wave,
we obtain:
\begin{equation}
  T_0 = {\left|\frac{\alpha_i}{\omega - \omega_0^i}\right|}^2.
  \label{eq:Lorentz}
\end{equation}

The complex band structure of the grating is shown in Fig.~\ref{fig:fig3}
for a height $h = 3~\mu{\rm{m}}$.
Fig.~\ref{fig:fig4} shows the zero-order
transmission spectra calculated for the same height
and two different angles of incidence.
They correspond to $\theta = 0^{\circ}$ (a) and $\theta = 12^{\circ}$ (b)
cuttings in Fig.~\ref{fig:fig3}(a).
The Lorentzian curves corresponding to Eq.\ (\ref{eq:Lorentz})
are added to the spectra. The constant $\alpha_i$ is fitted
to obtain the same maximum value as the spectra.
At normal incidence (Fig.~\ref{fig:fig4}(a)),
the broad resonances at 0.38 eV and 0.54 eV
(not represented in Fig.~\ref{fig:fig3})
are not separated enough to allow the comparison to
their respective Lorentzian functions to be made.
In such a case, the amplitude of the transmitted
wave is the sum of two complex Lorentzian functions,
and phase considerations make it more difficult
to fit to the spectra.

For the other peaks,
Fig.~\ref{fig:fig4} shows an excellent agreement
of the Lorentzian curves with the spectra.
Hence, the single-pole approximation is justified and
Eq.\ (\ref{eq:Lorentz}) gives the width at half maximum:
\begin{equation}
  \Delta \omega_0 = - \: 2 \: \Im(\omega_0^i).
  \label{eq:width}
\end{equation}
It justifies also the above definition of the quality factor
of the resonance, which corresponds to the more conventional
$Q = E/\Delta E$.

These transmission spectra allow some important
characteristics of the resonances to be pointed out.
At first, it must be noted that the flat band at about 0.35~eV
has no solution at normal incidence, and the peak $\gamma$
disappears at $\theta = 0^{\circ}$.
An energetic shift can be seen for
the horizontal surface-plasmon resonances between
$\theta = 0^{\circ}$ and $\theta = 12^{\circ}$,
whereas the vertical resonance at about 0.17~eV does not move.
Another point which comes out of the calculations
is that vertical resonances
in contiguous slits are nearly uncoupled.
As a consequence, a periodic structure is not necessary for them to exist:
the vertical surface-plasmon resonance could be excited in an unique slit,
and fully coupled with a few ones.
Hence, an important result is that the vertical resonance
can be efficiently excited by a strongly focalized beam.

In conclusion, we have carried out the calculation
of the complex dispersion curves of rectangular
metallic gratings in TM polarization.
CPBS are directly linked to experimental data, like transmission spectra.
The quality factor and the lifetime of resonances,
that they provide,
are crucial characteristics
in many applications already proposed
for the extraordinary effects appearing in
such transmission metallic gratings.
CPBS show pronounced discontinuities
located on Wood's anomalies, which can not
be evidenced in the usual energetic dispersion curves.
They allow to distinguish unambiguously
the two different natures of resonances:
\emph{horizontal and vertical surface-plasmon resonances}.
Similar results can be obtained for reflection gratings.
This work can also be extended in order to analyze resonances in more
complex geometries, as Ebbesen's structures \cite{Ebbesen98} for example,
for which a cutoff diameter exists for vertical modes.
The characteristics of the resonances studied in this Letter
open up new possibilities in photonic materials.
Moreover, these structures allow huge concentrations
of light in very small volumes,
which could have applications in fundamental optics \cite{Pendry99}
and in optoelectronics.

\begin{figure}
  \caption{Description of the metallic grating.
  Horizontal and vertical surface-plasmons are symbolically
  represented.}
  \label{fig:fig1}
\end{figure}
\begin{figure}[b!]
  \caption{
    Energetic (a) and quality-factor $Q$ (b) representation of the first four photonic bands
    of a rectangular metallic grating ($d = 3.5~\mu{\rm{m}}$, $a = 0.5~\mu\rm{m}$)
    with a height $h = 1.4~\mu{\rm{m}}$. The lower band is only computed for energies higher than 0.125 eV ($\lambda < 10~\mu{\rm{m}}$). Wood's anomalies (lines) are added to the energetic band structure (a).}
  \label{fig:fig2}
\end{figure}
\begin{figure}[b!]
  \caption{
    Energetic (a) and quality-factor $Q$ (b) representation of the first five photonic bands
    of a rectangular metallic grating ($d = 3.5~\mu{\rm{m}}$, $a = 0.5~\mu\rm{m}$)
    with a height $h = 3.0~\mu{\rm{m}}$. The lower band is only computed for energies higher than 0.125 eV ($\lambda < 10~\mu{\rm{m}}$). Wood's anomalies (lines) are added to the energetic band structure (a).}
  \label{fig:fig3}
\end{figure}
\begin{figure}[t!]
  \caption{
    Comparison between zero-order transmission spectra (solid lines) and Lorentzian curves (dotted lines) for a rectangular metallic grating ($d = 3.5~\mu{\rm{m}}$, $a = 0.5~\mu\rm{m}$, $h = 3.0~\mu{\rm{m}}$). The angle of incidence is (a) $\theta = 0^{\circ}$ and (b) $\theta = 12^{\circ}$.}
  \label{fig:fig4}
\end{figure}

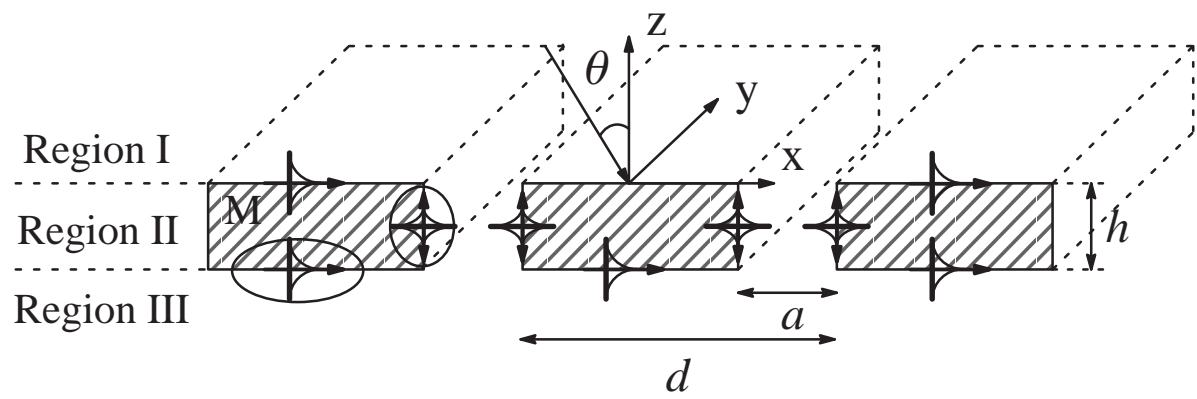
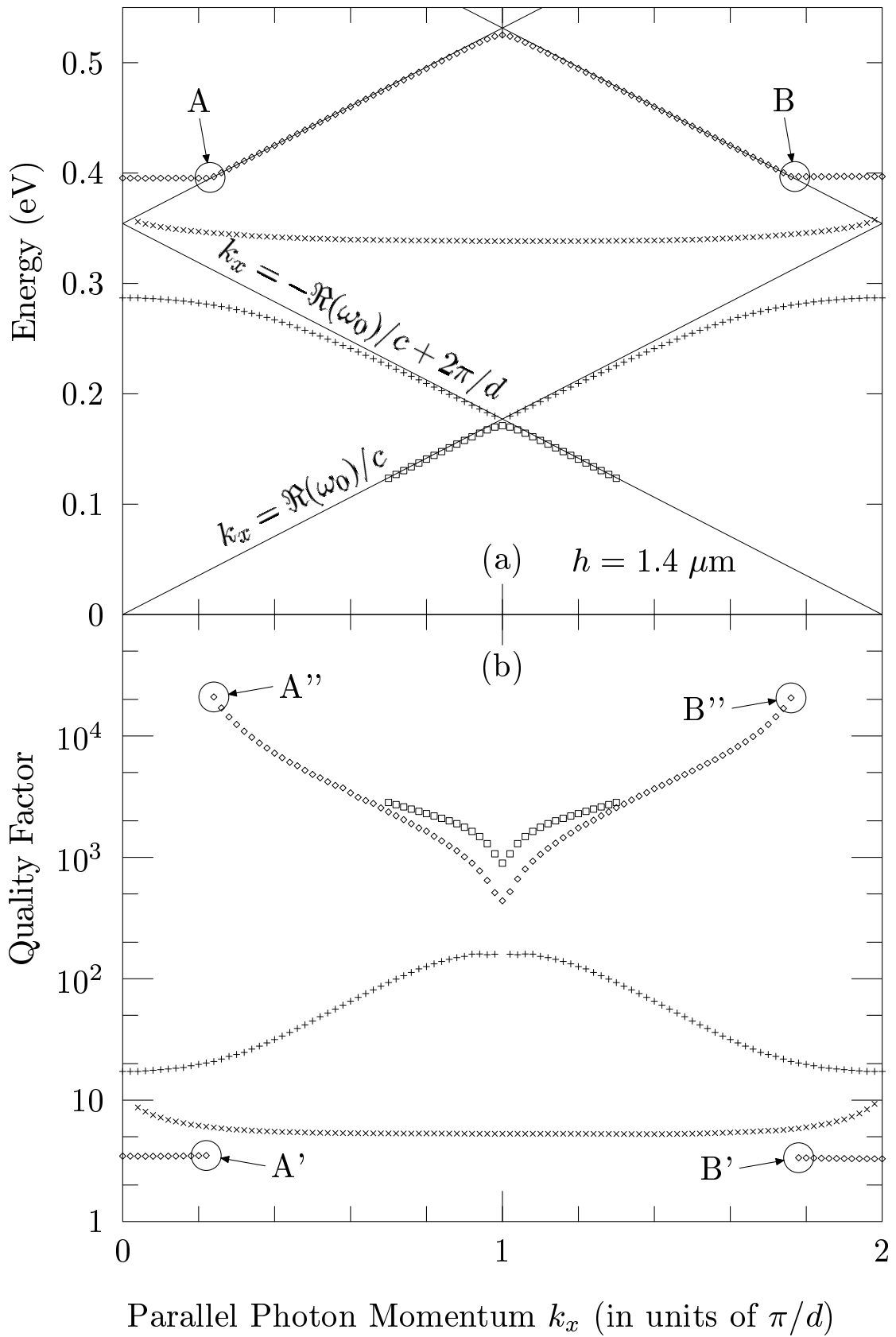
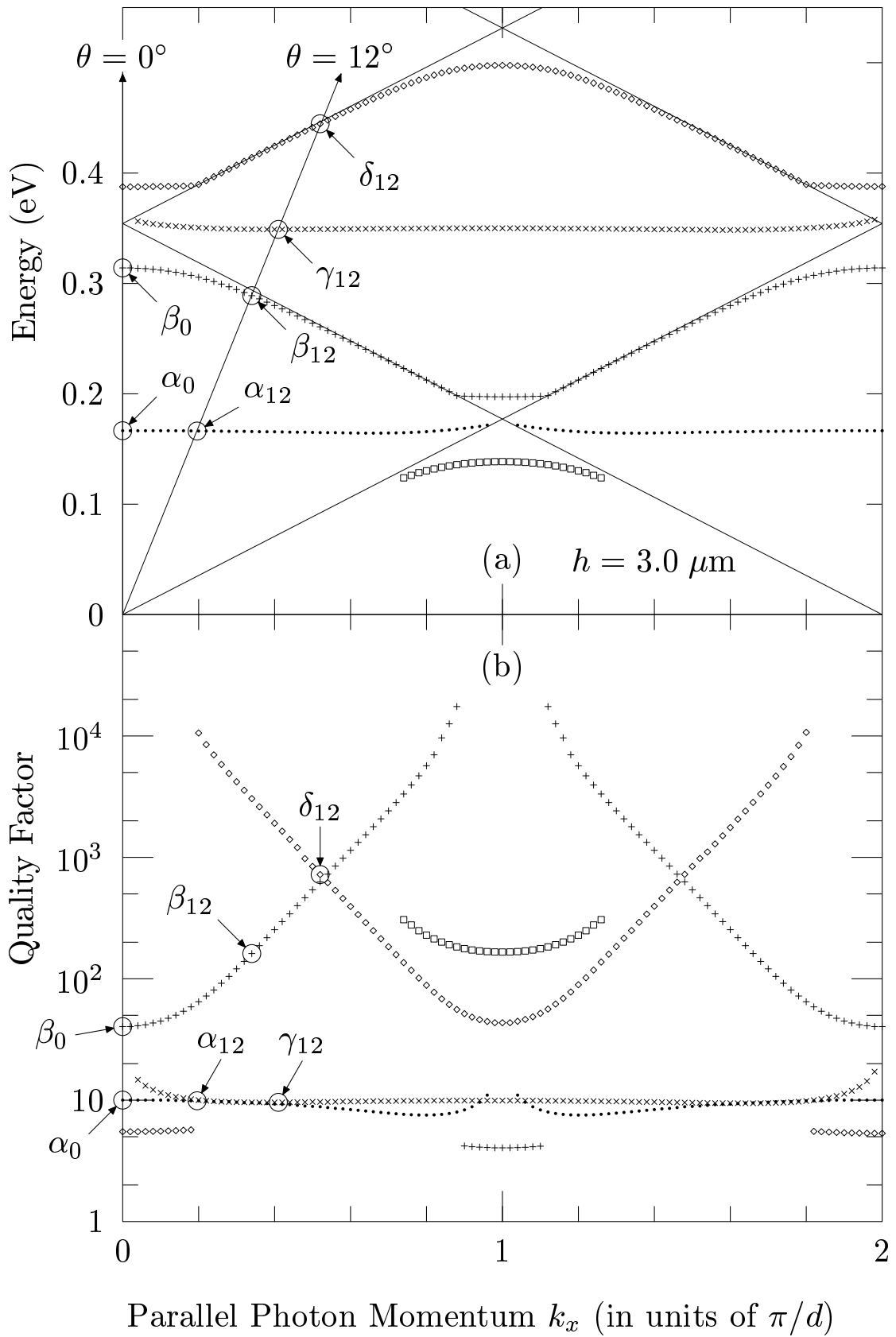
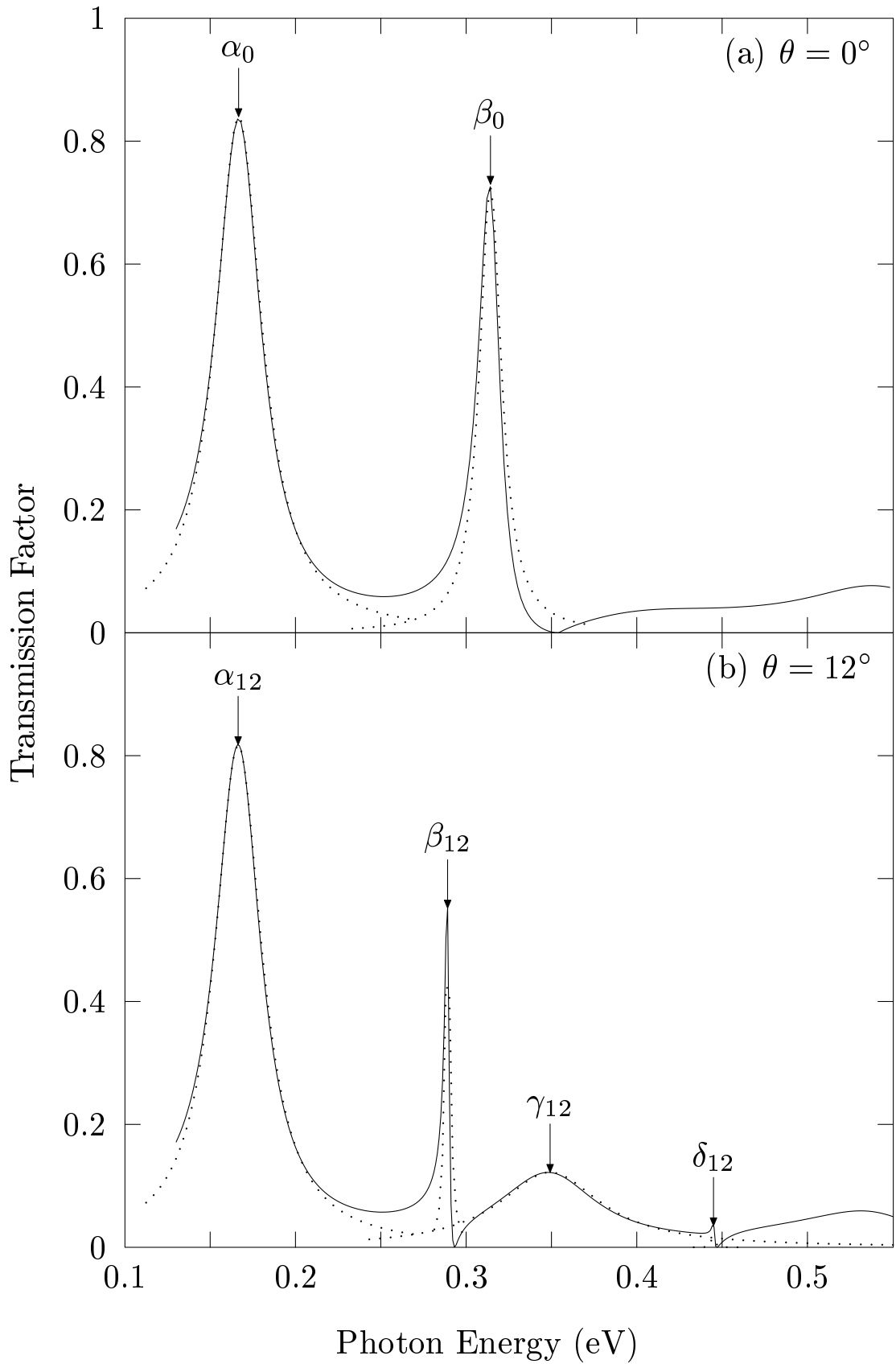
\end{document}